\documentclass[aps,showpacs,10pt,twocolumn]{revtex4}
\usepackage{epsf}

\begin{document}

\title{Sandwich test for Quantum Phase Estimation}         
\author{Avatar Tulsi\\
        {\small Department of Physics, Ganga Devi Mahila Mahavidyalaya (Women's College) \\ Patliputra University, Patna-800020, India}}
\email{tulsi9@gmail.com}

\begin{abstract}

Quantum Phase Estimation (QPE) has potential for a scientific revolution through numerous practical applications like finding better medicines, batteries, materials, catalysts etc. Many QPE algorithms use the Hadamard test to estimate $\langle \psi|U^{k}|\psi\rangle$ for a large integer $k$ for an efficiently preparable initial state $|\psi\rangle$ and an efficiently implementable unitary operator $U$. The Hadamard test is hard to implement because it requires controlled applications of $U^{k}$ which increase the circuit depth $T_{\rm max}$ by a factor of $\mathcal{O}(n)$ where $n$ is the system size. But the total run time $T_{\rm tot}$ of the Hadamard test scales only as $\mathcal{O}(k/\epsilon^{2})$ where $\epsilon$ is the desired accuracy of estimation. Recently, a Sequential Hadamard test (SHT) was proposed (arXiv:2506.18765) which requires controlled application of $U$ only, improving $T_{\rm max}$ by a factor of $\mathcal{O}(k)$. But the bottleneck is that $T_{\rm tot}$ of SHT scales as $\mathcal{O}(k^{3}\epsilon^{-2}r_{\rm min}^{-2})$ where $r_{\rm min}$ is the minimum value of $|\langle \psi|U^{k'}|\psi\rangle|$ among \emph{all} integers $k' \leq k$. Typically $r_{\rm min}$ is exponentially low and SHT becomes too slow.   

We present a new algorithm, the \rm{SANDWICH} test to address this bottleneck. Our algorithm uses efficient preparation of the initial state $|\psi\rangle$ to efficiently implement the \rm{SPROTIS} operator $R_{\psi}^{\phi}$ where SPROTIS stands for the Selective Phase Rotation of the Initial State. It sandwiches the SPROTIS operator between $U^{a}$ and $U^{b}$ for integers $\{a,b\} \leq k$ to estimate $\langle \psi|U^{k}|\psi\rangle$. The circuit depth of the Sandwich test is almost same as that of SHT. The total run time $T_{\rm tot}$ is $\mathcal{O}(k^{2}\epsilon^{-2} s_{\rm min}^{-6})$. Here $s_{\rm min}$ is the minimum value of $|\langle \psi|U^{\hat{k}}|\psi\rangle$ among all integers $\hat{k}$ which are values of the nodes of a random binary sum tree whose root node value is $k$ and leaf nodes' values are $1$ or $0$. It is difficult to analytically prove that $s_{\rm min} \not\ll 1$.But it can be reasonably expected in typical cases because there is extremely wide freedom in choosing the random binary sum tree. Numerical experiments are needed to confirm this.
\end{abstract} 

\pacs{03.67.Ac}

\maketitle

\section{Introduction}

Quantum Phase Estimation (QPE) is the holy grail of quantum simulation. Here the goal is to estimate the eigenvalues of a quantum system. In general, this is a hard problem~\cite{hadamard1,hadamard2,hard1,hard2,hard3,hard4,hard5}. But it is tractable in many cases of interest where we can efficiently prepare a quantum state with non-negligible overlap with the relevant eigenstates. It is widely believed that using QPE, a Quantum Computer (QC) can provide exponential advantage over the classical computers to find the eigenvalues but this belief is also questioned in~\cite{evidence1}. QPE has potential to bring a scientific revolution through numerous practical applications like discovery of better medicines, batteries, materials, catalysts etc~\cite{application1,application2,application3,application4,application5,application6,application7,application8,application9}. More details are provided in the review articles~\cite{review1,review2,review3}. 

There are several algorithms for QPE using various methods including Quantum Fourier Transform~\cite{textbook}, semiclassical Fourier transform~\cite{semiclassical}, variational methods~\cite{variational1,variational2,variational3}, Fourier-Filtering methods~\cite{fourier1,fourier2,fourier3,fourier4,fourier5}, matrix pencil methods~\cite{matrix1,matrix2}, quantum imaginary time evolution (QITE)~\cite{qite}, robust phase estimation~\cite{robust1,robust2,robust3,robust4,robust5}, Krylov-subspace methods~\cite{krylov1,krylov2}, and other important methods~\cite{other1,other2,other3,other4}. Apart from quantum simulation, QPE also has important applications in quantum algorithms~\cite{algorithm1,algorithm2,algorithm3,algorithm4,algorithm5,algorithm6,algorithm7}, quantum metrology~\cite{metrology1,metrology2,metrology3,metrology4}, quantum field theory~\cite{qft}, error mitigation~\cite{verifiedPE} etc.

Many algorithms for QPE use a simple quantum algorithm, the Hadamard test, to estimate the complex number $\langle \psi|U^{k}|\psi\rangle$ for a large integer $k$, an efficiently preparable initial state $|\psi\rangle$, and an efficiently implementable unitary operator $U$, whose eigenvalues need to be estimated through QPE. Typically, it is the time evolution operator corresponding to a time-independent Hamiltonian of a quantum system. 

The actual physical implementation of the Hadamard test is very hard because of the necessity of the $c \textendash U^{k}$ operator, the controlled application of $U^{k}$ using an ancilla qubit, also known as the \emph{control} qubit. Usually there is a constraint of locality on a Quantum Computer. It means that in one time step, we can implement only one local unitary operator which acts only on the neighboring qubits. The $c \textendash U^{k}$ operator is not a local operator because it acts on all qubits. The control qubit can be distributed onto a GHZ state of $O(n)$ ancilla qubits~\cite{ghz1,ghz2,ghz3,ghz4,ghz5,ghz6} so that each of the system's $n$ qubits has a neighboring ancilla qubit. But this will be very challenging as so much of ancilla qubits greatly adds to the decoherence of Quantum Computer. 

To precisely explain the locality constraint, let $T_{\rm{max}}(V)$ be the circuit depth required to implement any unitary operator $V$ on a Quantum Computer with the locality constraint. Then 
\begin{equation}
T_{\rm{max}}(U^{k}) = kT_{\rm{max}}(U). \label{TmaxkU}
\end{equation} 
We note that the efficient preparation of $|\psi\rangle$ means that we can efficiently implement a unitary operator $W$ which transforms $|0^{n}\rangle$ to $|\psi\rangle$ where $|0^{n}\rangle$ is the all-zero state in which all $n$ qubits are in $|0\rangle$ state. Typically 
\begin{equation}
T_{\rm{max}}(W) \ll T_{\rm{max}}(U) \ll T_{\rm{max}}(c \textendash U). \label{TWsmall}
\end{equation} 
So the circuit depth of the Hadamard test is
\begin{equation}
T_{\rm{max}}(\rm{Hadamard})  \approx T_{\rm{max}}(c \textendash U^{k}) = k T_{\rm{max}}(c \textendash U)  \label{tHadamardpre} 
\end{equation} 
To find $T_{\rm max}(c \textendash U)$, we note that $U$ can be decomposed into $L = T_{\rm max}(U)$ layers where each layer is a unitary operator $U_{l}$ $\left(l \in \{1,2,\ldots,L\}\right)$ of circuit depth $1$. So
\begin{equation}
U = \prod_{l = 1}^{L}U_{l},\ \ \ T_{\rm max}(U_{l}) = 1,\ \ \   l \in \{1,2,\ldots L\}.  \label{Uldefined}
\end{equation}
Each $U_{l}$ consists of local unitary operators which can be applied in parallel. The only way to locally implement $c \textendash U$ is to perform the following operations for each $U_{l}$: 
\begin{enumerate} 
\item We apply $\mathcal{O}(n)$ swaps to make the control qubit a neighbor of each of the $n$ qubits one-by-one. The control qubit has different neighbors after each swap. 
\item Between each of these swaps, we use the control qubit for a controlled application of the local unitary operator which acts on the neighboring qubits of the control qubit in $U_{l}$. 
\item Then we apply the swap operators again to bring back the control qubit to its original position. 
\end{enumerate} 
These steps cannot be applied in parallel. Hence the circuit depth of the controlled application of each layer is $\mathcal{O}(n)$ and the total circuit depth increases by the same factor. So Eq. (\ref{tHadamardpre}) can be rewritten as 
\begin{equation}
T_{\rm max}(\rm Hadamard) \approx k T_{\rm max}(c\textendash U) =\mathcal{O}(kn) T_{\rm max}(U). \label{tHadamard}
\end{equation} 
Without the locality constraint, it would have been only $kT_{\rm max}(U)$. Hence the circuit depth increases by the factor of $\mathcal{O}(n)$ because of the locality constraint. This is a huge increase in view of the fact that it is extremely difficult to develop a Quantum Computer of large depth because of decoherence. 

To address these bottlenecks, it is necessary to reduce the requirement of controlled operators. Several algorithms have been presented to do this but they have their own limitations. In~\cite{reference1,reference2}, it was shown that if we can prepare a superposition of $|\psi\rangle$ with an eigenstate of $U$ with a known eigenvalue then such controlled operations are not needed. But preparing such superpositions is itself highly susceptible to noise~\cite{ghz2,ghz4,ghz5,susceptible1,susceptible2,susceptible3}. Another recent approach~\cite{phaseretreival} is to use classical phase-retreival methods. But its performance has been demonstrated only numerically, not analytically. 

A better algorithm is presented in ~\cite{PRL_complex} using the Quantum Imaginary time evolution (QITE)~\cite{qite}. It requires $|\psi\rangle$ to have a short correlation length and $U$ to be a time evolution operator of a local Hamiltonian. Very recently, an algorithm of the Sequential Hadamard test (SHT) was presented in ~\cite{june25} to relax these requirements also. It requires controlled application of $U$ only, improving $T_{\rm max}$ by a factor of $\mathcal{O}(k)$. To estimate $\langle \psi|U^{k}|\psi\rangle$, SHT has to find and sum over the phase-differences between $\langle \psi|U^{k'-1}|\psi\rangle$ and $\langle \psi|U^{k'}|\psi\rangle$ for \emph{all} integers $k' \leq k$. The bottleneck is that the total run time $T_{\rm tot}$ of SHT scales as $\mathcal{O}(k^{3}/\epsilon^{2}r_{\rm min}^{2})$ where $\epsilon$ is the desired accuracy of estimation and $r_{\rm min}$ is minimum value of $|\langle \psi|U^{k'}|\psi\rangle|$ among \emph{all} integers $k' \leq k$. Typically $r_{\rm min}$ is exponentially low so SHT will become very slow. 

We point out that in original paper~\cite{june25}, SHT was presented in a different context where a unitary operator $U$ was written as a product of $N_{\rm gates}$ local gates $u_{l}$. In this context, they found the total sample complexity (total number of projective measurements) to be $\mathcal{O}(N_{\rm gates}^{2}\epsilon^{-2}r_{\rm min}^{-2})$. This can be easily generalized to the case of QPE where $U^{k}$ can be written as a product of $k$ operators $U$. The only difference is that in QPE, the total run time is not quantified in terms of the total number of measurements but the total number of required applications of $U$. In the $k'^{\rm th}$ iteration of SHT ($k' \leq k$), it will need $k'$ applications of $U$. Furthermore, SHT requires a total of $k$ iterations for all $k'$. This is why the total run time scales as $k^{3}$, not $k^{2}$ as it prima facie appears from $N_{\rm gates}^{2}$ dependance shown in the original paper~\cite{june25}.     

In this paper, we present a new quantum algorithm which we name as the \rm{SANDWICH} test. It sandwiches the \rm{SPROTIS} operator $R_{\psi}^{\phi}$ between two integer powers of $U$ to estimate $\langle \psi|U^{k}|\psi\rangle$. Here SPROTIS stands for the Selective Phase Rotation Of The Initial State $|\psi\rangle$. Mathematically, 
\begin{equation}
R_{\psi}^{\phi} = 1_{N}+\Phi|\psi\rangle\langle \psi| = WR_{0^{n}}^{\phi}W^{\dagger}, \label{SPROTIS}
\end{equation}  
where $1_{N}$ is the identity operator ($N = 2^{n}$), $R_{0^{n}}^{\phi}$ is the selective phase rotation of $|0^{n}\rangle$ and $\Phi$ is a complex number given by
\begin{equation}
 \Phi = e^{\imath 2\phi}-1 = 2\sin{\phi}e^{\imath \left(\phi+\pi/2\right)}. 
\end{equation}
The basic ingredient of the Sandwich test is the Sandwich operator $S$. For any two integer powers, $U^{a}$ and $U^{b}$, of unitary operator $U$, the Sandwich operator is given by $S = U^{a}R_{\psi}^{\phi}U^{b}$. This operator helps us to estimate $\langle \psi|U^{a+b}|\psi\rangle$ in terms of $\langle \psi|U^{b}|\psi\rangle$ and $\langle \psi|U^{b}|\psi\rangle$. This is then recursively used using a random binary sum tree to estimate $\langle \psi|U^{k}|\psi\rangle$.  

Like SHT, the Sandwich test also needs controlled application of only $U$, not for $U^{k}$ for $k > 1$ as required in the Hadamard test. The only additional operator in the Sandwich test is $R_{0^{n}}^{\phi}$ which is a multi-qubit controlled $2\times 2$ unitary gate. This can be efficiently implemented with a circuit depth of $\Theta(n)$ using the algorithms presented in ~\cite{mqcont1,mqcont2} if there is no locality constraint. But the locality constraint increases the depth by a factor of $\mathcal{O}(n)$ because of the similar reasons as given for Eq. (\ref{tHadamard}). Typically $T_{\rm max}(c\textendash U) \gg \mathcal{O}(n^{2})$ hence this overhead in the circuit depth is negligible. The spatial complexity of our algorithm is $n+1$ qubits where one ancilla is needed to do Hadamard test to estimate $\langle \psi|U|\psi\rangle$.

The total run time of the Sandwich test scales as $\mathcal{O}(k^{2} \epsilon^{-2} s_{\rm min}^{-6})$ where $s_{\rm min}$ is the minimum value of $|\langle \psi|U^{\hat{k}}|\psi\rangle|$ among all integers $\hat{k}$ which are values of the nodes of a random binary sum tree whose root node value is $k$. It is difficult to analytically prove that $s_{\rm min}$ is not very small. But it can be reasonably expected in typical cases because there is an extremely wide freedom in choosing the sum tree. Numerical experiments may be done to confirm this.

To give a plausible argument, we note that in Sandwich Test, we estimate $\langle \psi|U^{a+b}|\psi\rangle$ in terms of $\langle \psi|U^{a}|\psi\rangle$ and $\langle \psi|U^{b}|\psi\rangle$. This property helps us to \textit{jump} over bad $k'$'s for which $|\langle \psi|U^{k'}|\psi\rangle|$ is very small. This is unlike the SHT which has to cross through all integers $k' \leq k$. As an example, suppose the quantity $|\langle \psi|U^{k'}|\psi\rangle| \not \ll 1$ for $k' \in \{1,2,\ldots,100\}$ and we want to estimate $\langle \psi|U^{k}|\psi\rangle$ for $k =300$. Then if we are using SHT, a small value of $|\langle \psi|U^{k'}|\psi\rangle|$ for $k' = 101$ is enough to slow down the algorithm. But the Sandwich test allows us to \emph{jump} over the bad $k'$'s. In this example, even if all $|\langle \psi|U^{k'}|\psi\rangle|$'s for $k' \in \{101,102,\ldots,198\}$ are very small, it does not matter to us as we can use Sandwich test to estimate $\langle \psi|U^{k'}|\psi\rangle$ for $k' = 199$ by using our estimates of $\langle \psi|U^{k'}|\psi\rangle$ for $k' =99$ and $k' = 100$. 

We can also use multi-layer Sandwich tests if required. For example, we can apply two-layer Sandwich operators like $U^{a}R_{\psi}^{\phi_{1}}U^{b}R_{s}^{\phi_{2}}U^{c}$. It will help us to estimate $\langle \psi|U^{a+b+c}|\psi\rangle$ in terms of $\langle \psi|U^{a}|\psi\rangle$, $\langle \psi|U^{b}|\psi\rangle$, and $\langle \psi|U^{c}|\psi\rangle$. The angles $\phi_{1}$ and $\phi_{2}$ can be chosen to have multiple values to refine our estimate. In above example, it will help us to estimate $\langle \psi|U^{k'}|\psi\rangle$ for $k' = 199$ if the quantities $\langle \psi|U^{k'}|\psi\rangle \not \ll 1$ for $k' \in \{1,2,\ldots,66\}$. We can easily generalize it to higher-layer Sandwich operators. Higher-layer Sandwich operators can also be used to effectively shunt bad $k'$'s and to effectively increase $s_{\rm min}$. This is elaborated in Appendix. 

But what about the estimates of $\langle \psi |U^{k'}|\psi\rangle$'s for bad $k'$'s. We point out that even the Hadamard test will take a long time to estimate the arguments of such $\langle \psi |U^{k'}|\psi\rangle$'s. Fortunately, such $\langle \psi |U^{k'}|\psi\rangle$'s are not relevant for QPE as they do not contribute much to the Fourier convolution function which is estimated to an accuracy of $\mathcal{O}(\eta)$. Here $\eta$ is the minimum overlap of the initial state with the ground state of the system. It is assumed that $\eta \not \ll1$~\cite{fourier1,fourier2,fourier3,fourier4,fourier5}. So bad $k'$'s can be safely ignored. Thus, for practical purposes, $s_{\rm min}$ is actually the minimum value of $|\langle \psi|U^{\hat{k}}|\psi\rangle|$ among integer values of $\hat{k}$ of all nodes of our sum tree \emph{except} the root node. Because if $|\langle \psi|U^{k}|\psi\rangle|$ is very small for the root node value then we can safely ignore it for QPE. This is not true if $|\langle \psi|U^{\hat{k}}|\psi\rangle|$ is small for any other node.

Though we have presented the Sandwich test in the context of QPE, it can easily be generalized to the case considered in~\cite{june25} while presenting the Sequential Hadamard test. This is because the Sandwich test works if the SPROTIS operator is sandwiched between any two unitary operators, not just two integer powers of a particular unitary operator. We can also easily generalize the Sandwich test to continuous time setting.
   
In the context of QPE with an inexact (approximate) eigenstate, the importance of the SPROTIS operator was shown by the author long back in~\cite{mypaper}. There it was used to improve the spatial complexity of the Eigenpath Traversal Algorithm by Boixo, Knill, and Somma~\cite{BKS} with applications to quantum simulation and optimization. At that time, not much attention was given in the quantum computing community for QPE with inexact eigenstates. But now this topic has received wide attention. Hence the importance of the SPROTIS operator for this task needs to be understood in a better way as done in this paper. 

The paper is organized as following. The Sandwich test is presented in the next Section. Then we conclude with some discussion in Section III.

\section{The Sandwich Test}

The basic ingredient of the Sandwich test is the Sandwich operator $S$. For any two integer powers, $U^{a}$ and $U^{b}$, of unitary operator $U$, the Sandwich operator is given by 
\begin{equation}
S = U^{a}R_{\psi}^{\phi}U^{b}. \label{sandwichoperator}
\end{equation}
We define
\begin{eqnarray}
r_{a}e^{\imath \theta_{a}} &=& \langle \psi |U^{a}|\psi\rangle, \nonumber \\
r_{b}e^{\imath \theta_{b}} &=& \langle \psi |U^{b}|\psi\rangle, \nonumber \\
r_{a+b}e^{\imath \theta_{a+b}} &=& \langle \psi |U^{a+b}|\psi\rangle, \nonumber \\
s_{ab}& =& |\langle \psi |S|\psi\rangle|  \label{rthetadefined}\
\end{eqnarray}
The phase of $\langle \psi|S|\psi\rangle$ is irrelevant for the Sandwich test. The quantity $s_{ab}$ can be simplified using Eqs. (\ref{SPROTIS}-\ref{rthetadefined}) as
\begin{eqnarray}
s_{ab}  &=& \left|\langle \psi|U^{a}R_{\psi}^{\phi}U^{b}|\psi\rangle\right| \nonumber \\
           &=&\left| \langle \psi|U^{a}\left(1_{N}+\Phi |\psi\rangle\langle \psi|\right)U^{b}|\psi\rangle\right| \nonumber \\
           &=&\left|\langle \psi|U^{a+b}|\psi\rangle + \Phi \langle \psi|U^{a}|\psi\rangle \langle \psi|U^{b}|\psi\rangle\right| \nonumber \\
           &=&\left|r_{a+b}e^{\imath \theta_{a+b}} + \Phi r_{a}r_{b}e^{\imath\left(\theta_{a}+\theta_{b}\right)}\right| \nonumber \\
           &=&\left|r_{a+b} + 2r_{a}r_{b}\sin{\phi}e^{\imath \left(\theta_{a} + \theta_{b} -\theta_{a+b} + \phi+ \pi/2\right)}\right| \nonumber \\
          &=&\sqrt{r_{a+b}^{2}+4r_{a}^{2}r_{b}^{2} \sin^{2}{\phi}- 4 r_{a+b}r_{a}r_{b}\sin{\phi}\sin{\omega_{a+b}}}. \nonumber \\
& & \ 
\end{eqnarray}
So we have
\begin{equation}
\sin{\omega_{a+b}} = \frac{4r_{a}^{2}r_{b}^{2}\sin^{2}\phi + r_{a+b}^{2}- s_{ab}^{2}}{4r_{a+b}r_{a}r_{b}\sin{\phi}}. \label{sin_omega_ab}
\end{equation}
The quantity $\omega_{a+b}$ in above equations satisfies the following equation
\begin{equation}
\theta_{a+b} = \theta_{a}+\theta_{b} - \omega_{a+b} + \phi.  \label{theta_ab}
\end{equation}
Thus we can estimate $\theta_{a+b}$ using the estimates of the quantities $\{\theta_{a},\theta_{b},\omega_{a+b}\}$ as $\phi$ is already exactly known to us. Eq. (\ref{sin_omega_ab}) implies that estimates of the quantities $\{r_{a},r_{b},r_{a+b},s_{ab}\}$ can be used to get an estimate of $\sin \omega_{a+b}$. There can be two possible estimates of $\omega_{a+b}$ for a given estimate of $\sin \omega_{a+b}$. Two different values of $\phi$ can be used to resolve this ambiguity. 

Eq. (\ref{rthetadefined}) implies that the quantities $\{r_{a},r_{b},r_{a+b},s_{ab}\}$ can be estimated by repeatedly preparing $|\psi\rangle$, applying the corresponding unitary operator $\{U^{a},U^{b},U^{a+b},S\}$ on it and then averaging over the projective measurements of the resultant state onto $|\psi\rangle \langle \psi|$. These projective measurements are done by applying $W^{\dagger}$ on the resultant state and then measuring it in the computational basis of all $n$ qubits. We define  
\begin{equation}
s_{\rm min} = \min\{r_{a},r_{b},r_{a+b}\}.  \label{smin}
\end{equation}   
We can always choose $\phi$ to be $\Theta(1)$. Then Eq. (\ref{sin_omega_ab}) implies that $\Theta(m s_{\rm min}^{-6})$ projective measurements are required to get an estimate of $\omega_{a + b}$ with a variance of $1/m$.  

The Sandwich test can be used to estimate $\langle \psi |U^{k}|\psi\rangle$ by estimating its argument $\theta_{k}$ defined by
\begin{equation}
r_{k}e^{\imath \theta_{k}} = \langle \psi |U^{k}|\psi\rangle.
\end{equation}   
The modulus $r_{k}$ can always be estimated using projective measurements onto $|\psi\rangle\langle \psi|$ as explained earlier. To estimate $\theta_{k}$, let us consider a random binary sum tree. Let the nodes at height $h$ be denoted by indices $p_{h} \in \{1,2,\ldots,2^{h}\}$. Let the values of nodes be denoted by $\hat{k}_{h}^{p_{h}}$,  the subscript indicating the height of the node and the superscript indicating its position at that height. The value of the root node at height $0$ is chosen to be $\hat{k}_{0}^{1} = k$. We randomly choose numbers $x_{h}^{p_{h}}$ and $y_{h}^{p_{h}}$ so that
\begin{equation}
x_{h}^{p_{h}} + y_{h}^{p_{h}}= 1 \label{xy1}
\end{equation}
and
\begin{equation}
 0 < x_{\rm min} \leq  x_{h}^{p_{h}} \leq y_{h}^{p_{h}} < y_{\rm max}  = 1-x_{\rm min} < 1.  \label{xybounds}
\end{equation} 
The values of the children nodes are randomly chosen integers (including $0$) satisfying the following recursive relations
\begin{equation}
\hat{k}_{h+1}^{p_{h+1} = 2p_{h}-1} = \lceil x_{h}^{p_{h}}\hat{k}_{h}^{p_{h}} \rceil,\ \ \ \ \hat{k}_{h+1}^{p_{h+1} = 2p_{h}}  =  \lfloor y_{h}^{p_{h}} \hat{k}_{h}^{p_{h}} \rfloor  \label{nodevalues}
\end{equation}
Note that if the value of parent node is $1$ then one of its children node has value $1$ while other has value $0$. We choose the height of the tree, $h_{\rm max}$, to be the minimum value of $h$ for which values of all leaf nodes at height $h$ are either $0$ or $1$ but never more than $1$. We note that Eq. (\ref{xybounds}) imply that with each step, the value of node decreases by the minimum factor of $y_{\rm max}$ which can be chosen such that $y_{\rm max}^{h_{\rm max}}k$ is $1$ so $h_{\rm max}$ is $\Theta(\ln k)$. 

The unitary operator corresponding to a node is given by $\hat{k}_{h}^{p_{h}}$th power of $U$. We define
\begin{equation}
\hat{\theta}_{h}^{p_{h}} = \arg{\langle \psi |U^{\hat{k}_{h}^{p_{h}}}|\psi\rangle},
\end{equation}
We start with the leaf nodes at height $h_{\rm max}$ whose values are $0$ or $1$ and the corresponding unitary operators are $1_{N\times N}$ or $U$. We use the Hadamard test to estimate $\hat{\theta}_{h}^{p_{h}}$ for $h = h_{\rm max}$. This is $0$ or $\theta_{1} = \arg{\langle \psi|U|\psi\rangle}$. After getting this estimate, we use the Sandwich tests to estimate $\hat{\theta}_{h}^{p_{h}}$ for $h = h_{\rm max}-1$. Then we use these estimates to estimate $\hat{\theta}_{h}^{p_{h}}$ for $h = h_{\rm max}-2$. We continue this till we get to the root node $h=0$ which gives us an estimate of $\theta_{k} = \hat{\theta}_{0}^{p_{0}}$. Precisely, Eqs. (\ref{theta_ab}), (\ref{xy1}), and (\ref{nodevalues}) imply that  
\begin{equation}
\hat{\theta}_{h}^{p_{h}} = \theta_{h+1}^{p_{h+1} = 2p_{h}-1}+ \theta_{h+1}^{2p_{h}} - \hat{\omega}_{h}^{p_{h}} + \phi, 
\end{equation}
where $\hat{\omega}_{h}^{p_{h}}$ denotes the random variable obtained during the Sandwich test to estimate $\hat{\theta}_{h}^{p_{h}}$. 
The above recursive relation can be easily solved to get
\begin{equation}
\theta_{k} = \hat{\theta}_{0}^{1} = k\theta_{1} + \sum_{h =0}^{h_{\rm max}}\sum_{p_{h}=1}^{2^{h}}\hat{\omega}_{h}^{p_{h}}, \label{doublesum}
\end{equation}
where we have ignored the $\phi$-dependant terms as they are exactly known and do not contribute to the variance of random variable $\theta_{k}$. 

Note that $\hat{\omega}_{h}^{p_{h}}$ does not exist for trivial nodes who has at least one child node with value $0$. Because in this case the $\hat{\theta}$ value of a node is same as that of its non-zero value child node and no Sandwich test is needed to estimate it. Hence such trivial nodes are excluded from the double sum in above equation. Let $N_{h'}$ denote the total number of non-trivial nodes at height $h'$. The sum of values of all nodes at this height is equal to $k$. As a non-trivial parent node must have a minimum value of $2$, we have $N_{h'} =\Theta(k)$. 

Let us consider the variance of sum $\sum_{p_{h'}=1}^{2^{h'}}\hat{\omega}_{h'}^{p_{h'}}$ for a particular value $h = h'$. As each of $\hat{\omega}$'s are independent random variables, the variance of their sum is the sum of their variances. As mentioned after Eq. (~\ref{smin}), $\Theta(ms_{\rm min}^{-6})$ projective measurements are required to estimate $\hat{\omega}_{h'}^{p_{h'}}$'s with a variance of $1/m$. We do $\Theta(Ms_{\rm min}^{-6}/\left(\hat{k}_{h'}^{p_{h'}}\right)^{q})$ measurements to estimate $\hat{\omega}_{h'}^{p_{h'}}$ with a variance of $\left(\hat{k}_{h'}^{p_{h'}}\right)^{q}/M$. Choosing $M$ to be $\Theta(k^{q})$, we find the total variance to be 
\begin{equation}
\sum_{p_{h'}=1}^{2^{h'}}\left(\gamma_{h'}^{p_{h'}} \right)^{q},\ \  \gamma_{h'}^{p_{h'}} = \frac{\hat{k}_{h'}^{p_{h'}}}{k} \Longrightarrow \sum_{p_{h'}=1}^{2^{h'}}\gamma_{h'}^{p_{h'}}  =1. 
\end{equation}
because by the definition of sum tree, we have $\sum_{h'}\hat{k}_{h'}^{p_{h'}} =k$. Eq. (\ref{xybounds}) imply that $\max\{\gamma_{h'}^{p_{h'}} \} =y_{\rm max}^{h'}$. It is easy to show that the above variance is upper bounded by $y_{\rm max}^{h'(q-1)}$ for $q \geq 1$. Summing it over all $h'$, we find that the total variance is upper bounded by $(1-\gamma_{\rm max}^{q-1})^{-1} \leq 1$. 

Furthermore, for a particular node, we do $\Theta(Ms_{\rm min}^{-6}/\left(\hat{k}_{h'}^{p_{h'}}\right)^{q})$ projective measurements to get above variance. Each projective measurement requires $\Theta(\hat{k}_{h'}^{p_{h'}})$ applications of the unitary operator $U$ because $\hat{k}_{h'}^{p_{h'}}$ is the maximum integer power of $U$ which needs to be applied to estimate $\hat{\omega}_{h'}^{p_{h'}}$ using the Sandwich test. As $M$ is chosen to be $\Theta(k^{q})$ and as shown earlier, $N_{h'} = \Theta(k)$, the total number of required applications of $U$ for height $h'$ can be easily found to be 
\begin{equation}
 \Theta(k)s_{\rm min}^{-6}\sum_{p_{h'}=1}^{2^{h'}}\left(\gamma_{h'}^{p_{h'}} \right)^{1-q} = \Theta(k^{2})s_{\rm min}^{-6}x_{\rm min}^{h'(1-q)},
\end{equation}
where we have used Eq. (\ref{xybounds}) and the fact that $q \geq 1$. Summing it over all $h'$, we find the total run time complexity of the Sandwich test to estimate $\theta_{k}$ with a variance less than $1$ to be $\Theta (k^{2})s_{\rm min}^{-6}$. Obviously we have $k\theta_{1}$ term also in Eq. (\ref{doublesum}). But it can also be easily estimated by estimating $\theta_{1}$ to an accuracy of $\mathcal{1/k}$ using $\Theta(k^{2})$ Hadamard tests which requires same total run time also as each Hadamard test requires only one application of $U$. This contributes a constant-factor overhead to the total time complexity. 

If we want to estimate $\theta_{k}$ to an accuracy of $\epsilon$ then its variance should be $\epsilon^{2}$ and we need to do $\Theta(\epsilon^{-2})$ times more measurements. So the total run time complexity of our algorithm is 
\begin{equation}
T_{\rm tot} = \Theta\left(k^{2}\epsilon^{-2}s_{\rm min}^{-6}\right).
\end{equation}  
Thus the Sandwich test has a better run time than that of the Sequential Hadamard test. The $k^{2}$ dependance of the Sandwich test is of course better than $k^{3}$ dependance of the Sequential Hadamard test. But it is mostly the dependance on $r_{\rm min}$ and $s_{\rm min}$ which make the Sandwich test much better than the Sequential Hadamard test. Because typically $r_{\rm min}$ can be very small but $s_{\rm min}$ is not. It is difficult to prove it analytically but numerical experiments will be done to confirm this.

\section{Conclusion and Discussion}

We have presented a new quantum algorithm, the Sandwich test, to estimate $\langle s|U^{k}|s\rangle$ for an efficiently preparable initial state $|s\rangle$ and an efficiently implementable unitary operator $U$. The Sandwich test provides $\mathcal{O}(n)$ factor improvement of circuit depth complexity over the Hadamard test. It uses the SPROTIS operator which is not used by the Hadamard test. We have shown that the SPROTIS operator is a useful quantum resource for the important task of estimating $\langle s|U^{k}|s\rangle$. We only need to sandwich the SPROTIS operator between properly chosen unitary operators.

We also note that if we have a Quantum Computer of large enough circuit depth then we can use Hadamard test to estimate $\langle \psi|U^{k'}|\psi\rangle$ for $k' \geq 1$ also, not just for $k' = 1$ as discussed in the paper. Then we can use these values of $k'$ for the leaf nodes of our sum tree. It is easy to show that this will reduce the total run time by a factor of $\Theta ((k')^{2})$. 

\vspace{0.4cm}
\begin{widetext}
\section*{APPENDIX: MULTI-LAYER SANDWICH TESTS}

Consider the two-layer Sandwich operator $S = U^{a}R_{\psi}^{\phi_{1}}U^{b}R_{\psi}^{\phi_{2}}U^{c}$. Using the similar notation as done in the main text, we get the following expression
\begin{eqnarray}
s & = & |\langle \psi|U^{a}R_{\psi}^{\phi_{1}}U^{b}R_{\psi}^{\phi_{2}}U^{c}|\psi\rangle| \nonumber \\
  & = & |\langle \psi | U^{a} \left( 1_{N}+ \Phi_{1}|\psi\rangle \langle \psi| \right)  U^{b} \left( 1_{N}+ \Phi_{2}|\psi\rangle \langle \psi| \right)  U^{c} |\psi\rangle| \nonumber \\
 & = & |r_{a+b+c}e^{\imath \theta_{a+b+c}} + \Phi_{1}r_{a}r_{b+c}r^{\imath \left(\theta_{a} + \theta_{b+c}\right)} + \Phi_{2}r_{a+b}r_{c}e^{\imath \left(\theta_{a+b}+\theta_{c}\right)} + \Phi_{1}\Phi_{2}r_{a}r_{b}r_{c}e^{\imath \left(\theta_{a}+\theta_{b}+\theta_{c}\right)}| 
\end{eqnarray}
It is easy to see that if any of $r$'s is very small then it gets effectively shunted out from above equation. For example, if $r_{b+c} \ll 1$ then the second term in the above equation does not contribute much and we can get a good estimate of $\theta_{a+b+c}$ in terms of $\theta_{a}$, $\theta_{b}$, $\theta_{c}$ and $\theta_{a+b}$ provided none of $\{r_{a},r_{b},r_{c},r_{a+b}, r_{a+b+c}\}$ is too small. Also, $\Phi_{1}$ and $\Phi_{2}$ can be chosen to have different values to refine our estimates. Thus multi-layer Sandwich tests can be used to effectively increase the value of $s_{\rm min}$. Of course, exact analysis of higher-layer Sandwich operators is increasingly complicated. But if we are using a $t$-layer Sandwich operator then $s$ will be a sum of $2^{t}$ terms. Thus the degrees of freedom to estimate $\theta_{k}$ increases exponentially. Numerical experiments can be done to study higher-level Sandwich operators in typical cases. 

\end{widetext}

\end{document}